\documentclass[runningheads]{svmult}

\usepackage{makeidx}   % allows index generation
\usepackage{graphicx}  % standard LaTeX graphics tool
\usepackage{subeqnar}  % subnumbers individual equations
\usepackage{multicol}  % used for the two-column index
\usepackage{physprbb}  % modified textarea for proceedings,
\makeindex             % used for the subject index
\begin{document}
\title*{Forming Globular Cluster Systems in a Semi-analytic Scheme}
\toctitle{Forming Globular Cluster Systems  in a Semi-analytic Scheme}
%\protect\newline in the Particle Deflection Plane}
% allows explicit linebreak for the table of content
%
%
\titlerunning{GCs in semi-analytic models}
% allows abbreviation of title, if the full title is too long
% to fit in the running head
%
\author{Michael A. Beasley\inst{1}
\and Carlton M. Baugh\inst{2}
\and Duncan A. Forbes\inst{1}
\and Ray M. Sharples\inst{2}
\and Carlos S. Frenk\inst{2}}
\authorrunning{Mike Beasley et al.}
% if there are more than two authors,
% please abbreviate author list for running head
%
%
\institute{Astrophysics \& Supercomputing, Swinburne University, Hawthorn VIC 3122, Australia.
\and Department of Physics, University of Durham, Durham DH1 3LE, UK.}

\maketitle              % typesets the title of the contribution

\begin{abstract}
We apply the semi-analytical galaxy formation code of Cole et al. 
to investigate the formation of globular cluster (GC) systems in
hierarchical clustering scenarios. The nature of the model allows
us to investigate the properties of GC systems and their
parent galaxies within a cosmological framework, over a wide
dynamic range of mass and time resolution.
Assuming GCs form during mergers of gaseous systems, 
the metal-rich peak of the classical 'bimodal'
metallicity distribution of GCs naturally falls
out of our model, where such merging occurs over a wide range
of redshifts. The physical origin of old, metal-poor GCs (the
metal-poor peak) is harder to understand, since
their formation must be decoupled from the ongoing star formation
in these systems at high redshift ($z \sim$ 5). Within the
context of semi-analytic models in general, a possible solution 
lies in a cut-off in the GC formation efficiency at a 
characteristic local star formation rate.
\end{abstract}

\section{Introduction}

Globular cluster (GC) systems are generally regarded as 'probes of
galaxy formation'. They inhabit all luminous galaxies in the
local Universe, and a number of their properties (e.g. total
population, mean metallicity) scale with the mass of their host galaxy.
However, the potential for GC systems in revealing insights into the
formation of their hosts has been hampered by a lack of
understanding of the origin of GCs, and how GC formation 
relates to galaxy formation in general.
A number of models exist in the literature discussing the
formation of GC systems and their associated galaxies (or vice versa,
from a non-GC researcher viewpoint) through gaseous mergers 
(Ashman \& Zepf 1992; Barnes 1998), singular collapse (e.g Forbes, Brodie \&
Grillmair 1997) or the dissipationless accretion of satellites
(Forte, Martinez \& Muzzio 1982; Cote, Marzke \& West 1998).
Whilst all these models have had important successes in explaining the
observables (for example, the Ashman \& Zepf model 
predicted the observed colour bimodality seen in many GC systems), 
the distinction between the processes of rapid
collapse, merging and accretion is likely to be artificial in a 
Universe where we expect all these mechanisms to occur. 

A viable route to following the evolution of galaxies and their GC
systems within a cosmological framework is through 
semi-analytic modelling (e.g. White \& Rees 1978). We have applied the
fiducial semi-analytic model of Cole et al. (2000) to such 
a purpose, which is described in Beasley et al. (2002).

\subsection{A Synopsis of the Model}

\begin{figure}
\begin{center}
\includegraphics[width=0.55\textwidth]{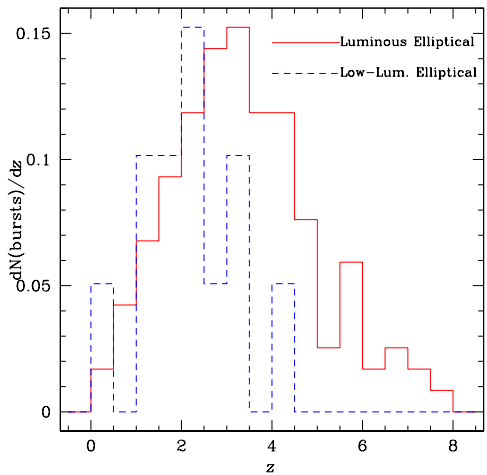}
\end{center}
\caption[]{Redshift distribution of major-mergers for a high-luminosity 
(M$_B$ - 5 log $h$ = --22.0, solid lines) and low-luminosity 
(M$_B$ - 5 log $h$ = --19.0, dashed lines) elliptical galaxy in the
SAM. Each merger induces a burst of star formation,
creating both field stars and GCs}
\label{merging}
\end{figure}

'Semi-analytic models' (SAMs) deal with the evolution of both baryonic and
non-baryonic matter using a number of analytical formulae and observed
scaling relations. The evolution of dark matter is either dealt with
using extended Press-Schechter formalism, or direct N-body numerical 
simulations. Gas, which is assumed to follow the distribution of dark
matter, is shock-heated when dark matter haloes merge. Over time, and
in the absence of further halo merging, this gas will cool and settle
into a disc\footnote{The taxonomy of these objects is ill defined,
  since they exhibit a range of physical sizes, gas-fractions and 
stellar populations. In the CDM picture, they may be classified as
Searle-Zinn fragments or galaxy building blocks.}
at the centre of the halo. Star formation is then allowed
to proceed in these discs in what is termed a 'quiescent' mode.
As dark matter haloes merge, they bring with them their
own stellar systems, the most massive of which becomes the central
galaxy and the others satellites of the new halo. 
If the dynamical timescales of these satellites are shorter 
than the timescale for halo merging, they will
merge. If this is a major merger, it will lead to a burst of star
formation.

\subsection{Metal-rich Globular Clusters}

A key aspect of our model are galaxy major mergers, which not only 
govern the morphology of the final galaxy (e.g. Baugh et al. 1996), 
but are also responsible for the formation the metal-rich
peak of the GC systems. We know that this latter assumption, 
one which was made by Ashman \& Zepf (1992), is at
least partially correct. Young, massive clusters have been identified in 
many nearby merging/interacting systems (e.g. Whitmore \& Schweizer 1995), 
some of which are expected to evolve into long-lived GCs 
(e.g Goudfrooij et al. 2001). 
In the SAM, each merger induces a burst of star 
formation, producing stars and GCs. 
The relative fraction of GCs formed in each burst
is determined by the GC formation efficiency, $\epsilon_{\rm GC}$
(the mass of GCs/mass of stars). 
This efficiency is determined from counting the
total populations of metal-rich GCs in present-day ellipticals, and
typically corresponds to $\sim$ 0.5\% efficiency.

The redshift distributions of major mergers for two galaxies in the
model are shown in Fig.~\ref{merging}. Each merger corresponds to
the formation of field stars and GCs. The high-luminosity elliptical
undergoes $\sim$ 80 mergers, the low-luminosity elliptical in
the figure has undergone $\sim$ 10. 
In this hierarchical picture, significant age
and metallicity sub-structure is present in this metal-rich
peak. Such sub-structure has now been detected
(e.g. Forbes et al. 2001; Kissler-patig et al., this volume),
and the question is now perhaps not {\it if} GCs are 
formed in mergers, but what {\it fraction} and {\it when}.

\subsection{Metal-poor Globular Clusters}

The source of metal-poor GCs (hereafter 'halo' GCs)
in the model are the star forming discs
which grow from gas which has cooled out of the dark matter halo.
Star formation proceeds as an 'accreting box' model, fresh gas
continually falls onto the discs, which is also expelled through
feedback from supernovae. The star formation rate in these 
discs is governed by the mass of gas available divided by an 
efficiency $\tau_*$. The fraction of these stars which
become GCs are again governed by $\epsilon_{\rm GC}$.
However, the drawback of this picture, where the progenitors of halo GCs
are gaseous galaxy 'building blocks' is that in our model, these
building blocks proceed to aggregate mass and form stars
quiescently until they reach metallicities at or above solar. 
This leads to metallicity distributions for the GCs which
are similar to that of the bulge stars.

A key to forming halo GCs from these building blocks is to
disconnect GC formation from the more general star formation
occurring at early epochs, and in Beasley et al. 
(2002) it is was necessary to truncate their formation at $z\sim$
5. A physically reasonable way of achieving this truncation is 
to tie $\epsilon_{\rm GC}$ to the local star formation rate (SFR)
(e.g. Larsen \& Richtler 2000). Hence, GCs are only allowed 
to form above a given SFR threshold, whereas field stars 
may form irrespective of this criterion. 
In the fiducial model of Cole et al., such a
SFR criterion fails for the halo GCs because the SFR peaks rather
late, yielding halo GCs which are too metal rich.

\begin{figure}
\begin{center}
\includegraphics[width=0.55\textwidth]{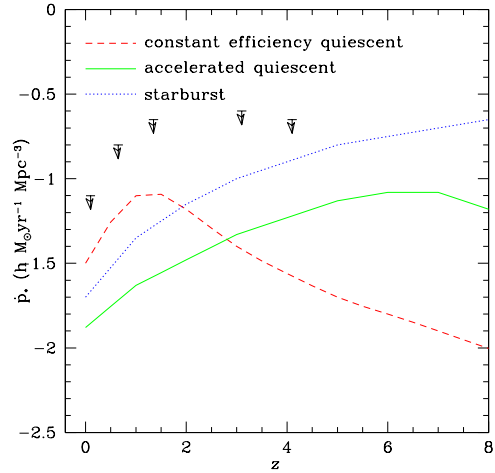}
\end{center}
\caption[]{Mean instantaneous star formation rates for 
three different assumptions about $\tau_*$. 
In the fiducial model of Cole et al. (2000), the star formation
history follows the quiescent lines. The upper limits represent the
observed 'maximal dust correction' from Somerville, Primack \&
Faber (2001)}
\label{sfr}
\end{figure}

\begin{figure}
\begin{center}
\includegraphics[width=0.6\textwidth]{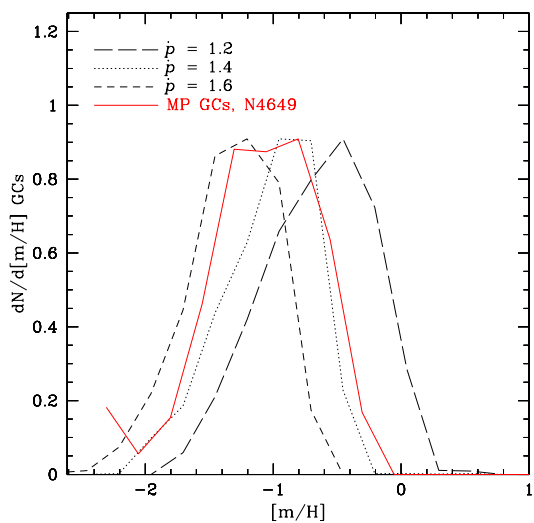}
\end{center}
\caption[]{Resultant metallicity distributions of metal-poor GCs
for different star formation rate thresholds in the SAM 
(broken lines). The solid line corresponds to the observed 
halo GCs in the Virgo
elliptical NGC~4649 (Forte et al., in preparation)}
\label{sfr1}
\end{figure}

We show the mean SFR in discs for three different assumptions 
about $\tau_*$ in Fig.~\ref{sfr}.
Nomenclature for the different star formation schemes comes from
Somerville, Primack \& Faber (2001), whose motivation was to
explain the observed characteristics of Lyman-break galaxies at
$z \geq 2$. Star formation schemes such as the 'accelerated
quiescent' mode (see Kauffmann, White \& Guideroni 1993; 
Somerville, Primack \& Faber 2001) which 
produce higher SFRs at early times do allow for a more natural
truncation of the halo GC formation.
This is illustrated in Fig.~\ref{sfr1}, where we show the
metallicity distributions for halo GCs adopting 
three different thresholds for the SFR in the accelerated
quiescent mode. This is compared to
Gemini/GMOS data for the halo GCs in the Virgo elliptical
NGC~4649 (Forte et al., in prep). For the SFRs shown
Fig.~\ref{sfr}, values of {\it \.{p}} = 1.3 adequately reproduce the
form of the halo GC metallicity distribution. The 
values in Fig.~\ref{sfr1} are illustrative only, since both the
exact form and normalisation of these SFRs are not strongly constrained
observationally (e.g. Steidel et al. 1999).

Although initially promising, how well such a scheme can be made 
to work in the broader context of galaxy formation and cosmology
(e.g. M. Santos, these proceedings) remains to be
seen. Importantly, the effects of other influences, such as the 
UV ionising background in SAMs
(e.g. Benson et al. 2002), and its relation to GC formation 
(e.g. Cen 2001) must be accounted for. 

\subsection{The GC-Spheroid Connection}

In our model, the entire metal-rich peak of the GC systems 
is created during merger-induced star
formation. The metal-poor peak reflects the mass spectrum and
accretion history of the galaxy progenitors. Since reproducing
the 'bimodal' metallicity distribution function (MDF) is regarded
as a key problem in GC research, we should ask the question how
well does the SAM reproduce the MDF of the spheroid stars?

\begin{figure}
\begin{center}
\includegraphics[width=0.6\textwidth]{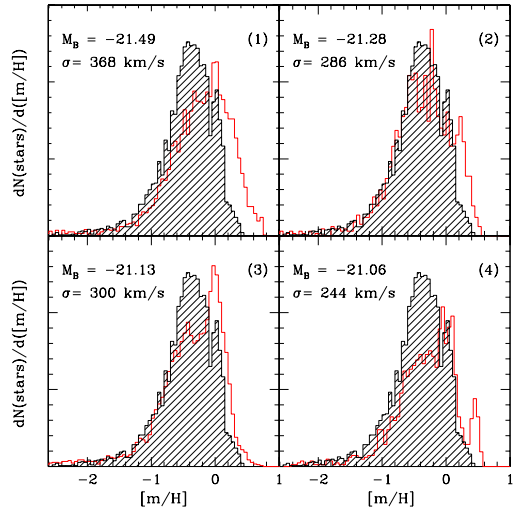}
\end{center}
\caption[]{Metallicity distribution functions for four model
ellipticals (open histograms) compared to the NGC~5128
distributions from Harris \& Harris (2002). In each panel,
M$_B$ refers to the B-band absolute magnitude of the model
galaxy, and $\sigma$ refers to its halo velocity dispersion}
\label{mdf}
\end{figure}

Until recently, this question was unimportant since {\it no} MDFs
of elliptical galaxies were observed. However, recent
observations of individual stars in the nearby elliptical
NGC~5128 (e.g. Harris \& Harris 2002), have made this an
important test.
Such a comparison is shown in Fig.~\ref{mdf} for four model
ellipticals compared to the stellar data of Harris \& Harris
(2002). The galaxies have been chosen to be similar to the luminosity
and from similar environments to NGC~5128, but otherwise have
been chosen arbitrarily. The agreement is surprisingly good,
considering that the models were not intended for making such
detailed comparisons.

We argue that such results demonstrate the utility of 
semi-analytic models in understanding the formation and 
evolution of GC systems in a cosmological
context.

\subsection{Acknowledgements}

MB thanks the organisers for an excellent workshop, and T. Puzia
in particular for the extracurricular activities. MB also thanks
Daisuke Kawata and Kenji Bekki for useful discussions.

%INDEX%%%%%%%%%%%%%%%%%%%%%%%%%%%%%%%%%%%%%%%%%%%%%%%%%%%%%%%%%%%%%%%
% Please check with the editor of your book whether he plans to
% include a "mutual" subject index - if so, please code your entries
% in the standard syntax. For your own purposes you may print your
% "personal" index by using the following commands:
%
%\clearpage
%\addcontentsline{toc}{section}{Index}
%\flushbottom
%\printindex
%%%%%%%%%%%%%%%%%%%%%%%%%%%%%%%%%%%%%%%%%%%%%%%%%%%%%%%%%%%%%%%%%%%%%

\end{document}